\documentclass[1p]{elsarticle}

\input{epsf}
\usepackage{amssymb}

 \def\frac#1#2{{\textstyle{{#1}\over
{#2}}}} 
\def\lsim{\mathrel{\rlap{\lower4pt\hbox{\hskip1pt$\sim$}}
\raise1pt\hbox{$<$}}}
\def\gsim{\mathrel{\rlap{\lower4pt\hbox{\hskip1pt$\sim$}}
\raise1pt\hbox{$>$}}} \def\sqr#1#2{{\vcenter{\vbox{\hrule
height.#2pt \hbox{\vrule width.#2pt height#1pt \kern#1pt \vrule
width.#2pt} \hrule height.#2pt}}}}

\def\beq{\begin{equation}} \def\eeq{\end{equation}}
\def\beqa{\begin{eqnarray}} \def\eeqa{\end{eqnarray}}

\long\def\symbolfootnote[#1]#2{\begingroup
\def\thefootnote{\fnsymbol{footnote}}\footnote[#1]{#2}\endgroup}

\journal{Physics Letters B}

\begin{document}

\begin{frontmatter}

\title{Modelling the reflective thermal contribution to the acceleration of the Pioneer spacecraft}

\author{F. Francisco\fnref{IPFN}}
\ead{frederico.francisco@ist.utl.pt}

\author{O. Bertolami\fnref{DFA,IPFN}}
\ead{orfeu.bertolami@fc.up.pt}
\ead[url]{http://web.ist.utl.pt/orfeu.bertolami}

\author{P. J. S. Gil\fnref{DEM}}
\ead{p.gil@dem.ist.utl.pt}

\author{J. P\'aramos\fnref{IPFN}}
\ead{http://paramos@ist.edu}
\ead[url]{http://web.ist.utl.pt/jorge.paramos}

\address[IPFN]{Instituto de Plasmas e Fus\~ao Nuclear\\Instituto Superior T\'ecnico, Universidade T\'ecnica de Lisboa\\ Av. Rovisco Pais 1, 1049-001 Lisboa, Portugal\\~}

\address[DFA]{Departamento de F\'{\i}sica e Astronomia\\Faculdade de Ci\^encias, Universidade do Porto\\Rua do Campo Alegre 687, 4169-007 , Porto, Portugal\\~}

\address[DEM]{Departamento de Engenharia Mec\^anica and IDMEC - Instituto de Engenharia Mec\^anica\\Instituto Superior T\'ecnico, Universidade T\'ecnica de Lisboa\\Av.\ Rovisco Pais 1, 1049-001 Lisboa, Portugal\\~}

\date{\today}


\begin{abstract}
	
We present an improved method to compute the radiative momentum transfer in the Pioneer 10 \& 11 spacecraft that takes into account both diffusive and specular reflection. The method allows for more reliable results regarding the thermal acceleration of the deep-space probes, confirming previous findings. A parametric analysis is performed in order to set an upper and lower-bound for the thermal acceleration and its evolution with time.

\end{abstract}

\end{frontmatter}


\section{Introduction}

\subsection{General Background}

For over a decade, the Pioneer anomaly has stood out as an open question in physics. The existence of this apparently constant sun-bound acceleration on the Pioneer 10 and 11 deep-space probes was first raised in 1998 by a team of scientists at the Jet Propulsion Laboratory (JPL) \cite{Anderson1998}. In the subsequent years, this anomalous acceleration was studied in detail, with the latest results pointing towards a constant value of $(8.74 \pm 1.33) \times 10^{-10} ~ \rm{m/s^2}$ \cite{Anderson2002}.

The existence of this effect has been independently confirmed through alternative data analyses \cite{Markwardt2002,Levy2009,Toth2009}. At least two of these analyses do also allow for a non-constant anomalous acceleration \cite{Markwardt2002,Toth2009}.

Throughout the last decade, numerous attempts have been made to explain the Pioneer anomaly. These can be mainly divided into two categories: conventional \cite{Katz1999, Scheffer2003} and new physics explanations \cite{Bertolami2004, Reynaud2005, Moffat2006, Bertolami2007}. It has also been shown that the Kuiper Belt is not the cause of the anomalous acceleration, considering several models for its mass distribution \cite{Bertolami2006}.

	
\subsection{Thermal Effects}

The initial assessment of systematic effects made in Ref. \cite{Anderson2002} asserted that any acceleration arising from thermal dissipation would be too small to account for the Pioneer anomaly. However, early in this discussion, some argued that thermal effects could provide a viable explanation for the detected anomaly. Indeed, two works argued, albeit on a qualitative basis, that there was sufficient on-board thermal energy to account for the anomalous acceleration \cite{Katz1999, Scheffer2003}.

Some time later, the effort to obtain a quantitative description of the effects of the thermal emissions of the spacecraft gained momentum, with three independent studies underway for the last few years. The first results were published by our team in 2008 \cite{Bertolami2008,Bertolami2010}. The estimate was performed using a method based in a distribution of point-like Lambertian and isotropic radiation sources. A set of test cases carried out in the study showed that this method can effectively model the main contributions at work, while keeping the desired simplicity, computational flexibility and speed \cite{Bertolami2008,Bertolami2010}.

The results indicated that between 33\% and 67\% of the observed acceleration can be explained by the thermal emissions of the spacecraft itself \cite{Bertolami2008,Bertolami2010}. This was confirmed by early figures produced by the team at ZARM \cite{Rievers2010}. It has also been reported that an analysis is underway by the JPL based team \cite{Toth2008}.

This paper builds on the previous work and presents a comprehensive analysis based on a direct modeling of reflection, in opposition to estimates based on surface reflectivity, as obtained in Ref.\ \cite{Bertolami2008}. In addition, a parametric analysis is carried out in order to establish reliable bounds for the obtained results. These have been confirmed by the analysis of telemetry data and finite-element, detailed modelling of the Pioneer's probes performed by the ZARM group \cite{Rievers2011}.


\section{Point-like Source Method}


\subsection{Motivation}

As outlined in Ref.\ \cite{Bertolami2008}, we adopted an approach that maintains a high degree of computational flexibility and speed, allowing for analysis of different scenarios and contributions. This approach is motivated by the limitations inherent to the characterization of the anomalous acceleration itself. Indeed, one of the analysis of the flight data shows that both a constant acceleration and one with a linear decay of a period greater than 50 years are compatible with the data \cite{Markwardt2002}.

In another alternative determination, while testing for the constancy of the acceleration, a so-called ``jerk term'' is found to be consistent with the expected temporal variation of a recoil force due to heat generated on board \cite{Toth2009}. This is essential if the hypothesis of a thermal origin for the Pioneer anomaly is to be considered, as such source would inevitably lead to a decay with at least the same rate as the power available onboard. Possible causes for an enhanced decay include {\it e.g.} degradation of thermo-couples, stepwise shutdown of some systems and instruments, {\it etc}. \cite{Markwardt2002}.

Bearing all this in mind, our method was designed to keep all the physical features of the problem visible and all steps easy to scrutinize. Although it can be argued that this simplicity and transparency was achieved at the expense of the accuracy of the method, a series of test cases were performed to demonstrate the robustness of the results \cite{Bertolami2008, Bertolami2010}. These test cases validate our approach, as they show that, for reasonable assumptions, the possible lack of accuracy caused by our modeling approach is much smaller than the accuracy in the characterization of the acceleration itself.

In this paper, the usefulness of these features of the method is further put to the test, as a parametric analysis of the problem is performed in order to sort out the relative importance of the different parameters involved.


\subsection{Radiative Momentum Transfer}\label{sec:radmomtransf}

Our method is based on a distribution of point-like radiation sources that models the thermal radiation emissions of the spacecraft.

All the subsequent formulation of emission and reflection is made in terms of the Poynting vector-field. We thus begin with the vector-field descriptions for the radiation emitting surfaces, modeled as Lambertian sources. The time-averaged Poynting vector field for a Lambertian source located at $\mathbf{x}_0$ is given by

\begin{equation}
	\label{lambertian}
	\mathbf{S}(\mathbf{x})={W \cos \theta \over \pi ||\mathbf{x}-\mathbf{x}_0||^2}{\mathbf{x}-\mathbf{x}_0 \over ||\mathbf{x}-\mathbf{x}_0||},
\end{equation}

\noindent where $W$ is the emissive power and $\theta$ is the angle with the surface normal. The value of $\cos \theta$ can be replaced by the inner product between the unitary emitting surface normal $\mathbf{n}$ and the emitted ray vector $(\mathbf{x}-\mathbf{x}_0)$ divided by its norm, allowing us to rewrite Eq.~(\ref{lambertian}) in a more practical form:

\begin{equation}
	\label{lambertian2}
	\mathbf{S}(\mathbf{x})={W \over \pi ||\mathbf{x}-\mathbf{x}_0||^2} \left( \mathbf{n} \cdot {\mathbf{x}-\mathbf{x}_0 \over ||\mathbf{x}-\mathbf{x}_0||} \right) {\mathbf{x}-\mathbf{x}_0 \over ||\mathbf{x}-\mathbf{x}_0||}.
\end{equation}

The amount of energy illuminating a given surface $E_{\rm ilum}$ can be obtained by computing the Poynting-vector flux through the illuminated surface:

\begin{equation}
	E_{\rm ilum} = \int \mathbf{S} \cdot \mathbf{n}_{\rm ilum}~ dA,
\end{equation}

\noindent where $\mathbf{n}_{\rm ilum}$ is the normal vector of the illuminated surface.

The thermal radiation (infrared radiation) illuminating a surface will yield a force on that surface. This force per unit of area is the \emph{radiation pressure} $p_{\rm rad}$, given by

\begin{equation}
	p_{\rm rad}={\mathbf{S} \cdot \mathbf{n}_{\rm ilum} \over c},
\end{equation}

\noindent that is, the energy flux divided by the speed of light. This result should be multiplied by a factor $\alpha$, that varies between $1$ for full absorption and $2$ for full reflection, which allows for an estimate of the reflection (as assessed in Refs. \cite{Bertolami2008,Bertolami2010}). However, a more rigorous treatment of reflection is presented in the next sections.

Integrating the radiation pressure on a surface, we obtain the exerted force

\begin{equation}
	\label{force_integration}
	\mathbf{F} = \int {\mathbf{S} \cdot \mathbf{n}_{\rm ilum} \over c} {\mathbf{S} \over ||\mathbf{S}||} dA.
\end{equation}

\noindent The interpretation of this integration is not always be straightforward: to obtain the force exerted by the radiation on the emitting surface, the integral should be taken over a closed surface encompassing the latter. Analogously, the force exerted by the radiation on an illuminated surface requires an integration surface that encompasses it.

Furthermore, considering a set of emitting and illuminated surfaces implies the proper account of the effect of the shadows cast by the various surfaces, which is then subtracted from the estimated force on the emitting surface. One may then straightforwardly read the thermally induced acceleration,

\begin{equation}
	\mathbf{a}_{\rm th}={\sum_i \mathbf{F}_i \over m_{\rm pio}}.
\end{equation}


\subsection{Reflection Modeling -- Phong Shading}\label{sec:phong}

In this study a more accurate modeling of reflection is carried out. The geometric configuration of the Pioneer 10 and 11 probes is bound to cause several reflections that may potentially weigh on the final result. We use a method known as \emph{Phong Shading}, a set of techniques and algorithms commonly used to render the illumination of surfaces in 3D computer graphics. It was developed in the 1970's by Bui Tuong Phong at the University of Utah and published in his Ph. D. thesis \cite{Phong}.

This method comprises two distinct components:

\begin{itemize}
	\item a reflection model including diffusive and specular reflection, known as \emph{Phong reflection model};
	\item an interpolation method for curved surfaces modeled as polygons, known as \emph{Phong interpolation}.
\end{itemize}

The Phong reflection model is based on an empirical formula that gives the illumination value of a given point in a surface $I_p$ as

\begin{equation}
	I_p=k_a i_a + \sum_{m \in \mathrm{lights}} \left[k_d (\mathbf{l}_m \cdot \mathbf{n})i_d + k_s (\mathbf{r}_m \cdot \mathbf{v})^{\alpha} i_s \right],
\end{equation}

\noindent where $k_a$, $k_d$ and $k_s$ are the ambient, diffusive and specular reflection constants, $i_a$, $i_d$ and $i_s$ are the respective light source intensities, $\mathbf{l}_m$ is the direction of the light source $m$, $\mathbf{n}$ is the surface normal, $\mathbf{r}_m$ is the direction of the reflected ray, $\mathbf{v}$ is the direction of the observer and $\alpha$ is a ``shininess'' constant (the larger it is, the more mirror-like the surface is).

This method provides a simple and straightforward way of modeling the various components of reflection, as well as a more accurate accounting of the thermal radiation exchanges between the surfaces on the Pioneer spacecraft. In principle, there is no difference between the treatment of infrared radiation, in which we are interested, and visible light, for which the method was originally designed (allowing for a natural wavelength dependence of the above material constants).

Given the presentation of the thermal radiation put forward in subsection \ref{sec:radmomtransf}, the Phong shading methodology was adapted from a formulation based on \emph{intensities} (energy per surface unit per surface unit of the projected emitting surface) to one based on the energy-flux per surface unit (the Poynting vector).


\subsection{Computation of Reflection}

Using the formulation outlined in section \ref{sec:phong}, we separately compute the diffusive and specular components of reflection in terms of the Poynting vector-field. We begin by writing the reflected radiation Poynting vector-field for the diffusive component of the reflection as

\begin{equation}
	\label{diffusive_reflection}
	\mathbf{S}_{\mathrm{rd}}(\mathbf{x},\mathbf{x}')={k_d |\mathbf{S}(\mathbf{x}')\cdot \mathbf{n}| \over \pi ||\mathbf{x}-\mathbf{x}'||^2} (\mathbf{n} \cdot (\mathbf{x}-\mathbf{x}')) {\mathbf{x}-\mathbf{x}' \over ||\mathbf{x}-\mathbf{x}'||},
\end{equation}

\noindent while the specular component reads

\begin{equation}
	\label{specular_reflection}
	\mathbf{S}_{\mathrm{rs}}(\mathbf{x},\mathbf{x}')={k_s |\mathbf{S}(\mathbf{x}')\cdot \mathbf{n}| \over {2 \pi \over 1+ \alpha} ||\mathbf{x}-\mathbf{x}'||^2} (\mathbf{r} \cdot (\mathbf{x}-\mathbf{x}'))^{\alpha} {\mathbf{x}-\mathbf{x}' \over ||\mathbf{x}-\mathbf{x}'||}.
\end{equation}

\noindent In both cases, the reflected radiation field depends on the incident radiation field $\mathbf{S}(\mathbf{x}')$ ($\mathbf{x}'$ is a point on the reflecting surface) and on the reflection coefficients $k_d$ and $k_s$, respectively. Using Eqs.~(\ref{diffusive_reflection}) and (\ref{specular_reflection}), we can compute the reflected radiation field by adding up these diffusive and specular components. From the emitted and reflected radiation vector-fields, the irradiation of each surface is computed and, from it, a calculation of the force can be performed through Eq.~(\ref{force_integration}). This formulation allows for the determination of the force on the whole spacecraft, accounting for radiation that is reflected and absorbed by the various surfaces, as well as that which is propagated into space.

In the modeling of the actual spacecraft, once the radiation source distribution is put into place, the first step is to compute the emitted radiation field and the respective force exerted on the emitting surfaces. This is followed by the determination of which surfaces are illuminated and the computation of the force exerted on those surfaces by the radiation. At this stage, we get a figure for the thermal force without reflections. The reflection radiation field is then computed for each surface and subject to the same steps as the initially emitted radiation field, leading to a determination of thermal force with one reflection.

This method can, in principle, be iteratively extended to as many reflection steps as desired, considering the numerical integration algorithms and available computational power. If deemed necessary, each step can be simplified through a discretization of the reflecting surface into point-like reflectors.


\section{Pioneer Thermal Model}


\subsection{Model Features}

The first step in the pursuit of a reliable estimate of the thermal effects is to build a geometric model of the Pioneer spacecraft. Bearing in mind the approach outlined in the previous sections and Ref. \cite{Bertolami2008}, we aim for a suitable balance between detail and simplicity.

The geometric model used in this study retains the most important features of the Pioneer spacecraft, namely:

\begin{itemize}
	\item the parabolic high-gain antenna,
	\item the main equipment compartment behind the antenna,
	\item two radio-thermal generators (RTGs), cylindrical in shape, each connected to the main compartment through a truss.
\end{itemize}

\begin{figure*}
	\begin{center}
	\epsfxsize=\columnwidth 
	\epsffile{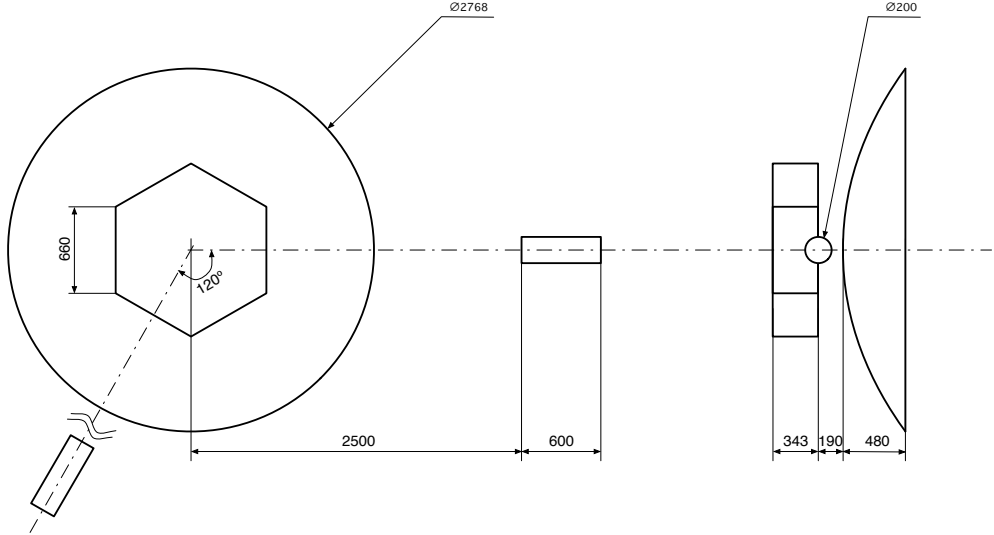} 
	\caption{Schematics of the Pioneer geometric model used in our study, with relevant dimensions (in $\rm{mm}$); second RTG truss is not represented to scale. Lateral view indicates the relative position of the RTGs, box compartment and the gap between the latter and the high-gain antenna.}
	\label{pioneer_schematic}
	\end{center}
\end{figure*}

\noindent The full shape and dimensions of the geometric model are depicted in Fig. \ref{pioneer_schematic}.

This model simplifies the surface features and minor details of the spacecraft. It has been tested through specific test-cases, as presented in Ref. \cite{Bertolami2008,Bertolami2010}, which show that its effect on the final result can be safely ignored for the purposes of this study.

The modeling of the thermal radiation emissions is constructed through a distribution of a small number of carefully placed point-like sources. This source distribution should reflect the real emissions of the spacecraft as closely as possible.

It is important to highlight that the spin-stabilization of the Pioneer probes considerably simplifies the task of modeling the force generated from thermal emissions: the effect of all radial emissions is cancelled out after each complete revolution of the spacecraft. The only remaining contribution is along the antenna's axis (here taken as the $z$-axis).

When considering thermal radiation sources, two main components of the probe can be identified:

\begin{itemize}
	\item the RTGs, where the main power source of the spacecraft is located,
	\item the main equipment compartment, where the majority of the power is consumed.
\end{itemize}

\begin{figure}
	\begin{center}
	\epsfxsize=0.57\columnwidth
	\epsffile{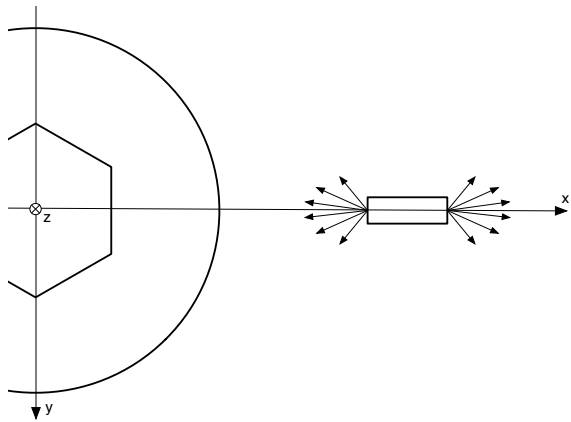} 
	\caption{Schematics of the two Lambertian sources used to model each RTG.}
	\label{RTG_sources}
	\end{center}
\end{figure}

The RTGs can be easily and effectively modeled by two Lambertian sources, one at each base of the cylinder, as shown in Fig.~\ref{RTG_sources}. The radiation from the source facing outwards will radiate directly into space in a radial direction and its contribution will cancel-out. However, the radiation emitted towards the centre of the spacecraft will be reflected by both the high-gain antenna and the main equipment compartment.

The analysis of the main equipment compartment is divided between the front, back and lateral walls.

The front wall of the latter (facing away from the sun and where the heat-dissipating louvers) will emit radiation directly into space, not illuminating any other surface. It can then be modeled through a single radiation source, without impact on the final result.

\begin{figure}
	\begin{center}
	\epsfxsize=0.47\columnwidth
	\epsffile{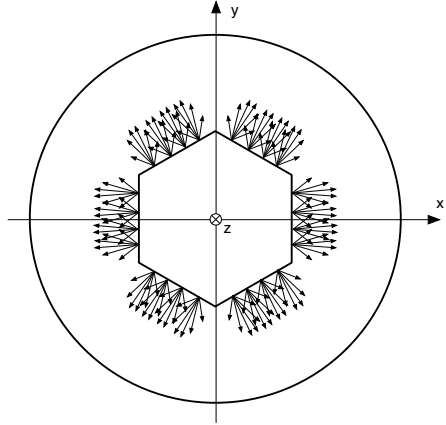} 
	\caption{Schematics of the configuration of Lambertian sources used to model the lateral walls of the main equipment compartment.}
	\label{equipment_sources}
	\end{center}
\end{figure}

The side walls of this compartment are each modeled by four Lambertian sources, as seen in Fig.~\ref{equipment_sources}. A previously conducted convergence analysis shows that this provides a reasonable degree of accuracy \cite{Bertolami2008,Bertolami2010}. This radiation will reflect mainly on the high-gain antenna.

\begin{figure}
	\begin{center}
	\epsfxsize=0.21\columnwidth
	\epsffile{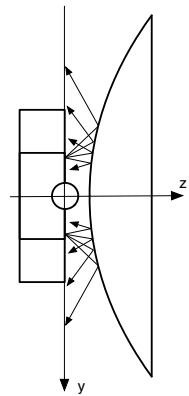} 
	\caption{Schematics of the configuration of Lambertian sources used to model the back wall of the main equipment compartment and the first reflection on the main antenna dish.}
	\label{back_wall}
	\end{center}
\end{figure}

Our previous results lacked the contribution of the back wall of the main equipment compartment (facing the high-gain antenna). The radiation from this wall will, in a first iteration, reflect off the antenna and add a contribution to the force in the direction of the sun, as depicted in Fig~\ref{back_wall}. This back wall was modeled using a set of six Lambertian sources evenly distributed in the hexagonal shape.

The relevant contributions for this analysis can be summarized in Table \ref{pioneer_components}, with each of them indexed for reference in the following sections.

\begin{table*}
	\begin{center}
	\caption{Indexing of the components considered in this study}
	\vskip 0.2cm
	\begin{tabular}{c c c}
	
		Emitting surface					& Reflecting surface					& Index \\
		\hline
		\hline
		Side of main compartment	& High-gain antenna dish				& 1.1 \\
		RTG									& High-gain antenna dish				& 2.1 \\
		RTG									& Side of main compartment	& 2.2 \\
		Back of main compartment	& High-gain antenna dish				& 3.1 \\
		Front of main compartment	& None									& 4 
		\label{pioneer_components}
	\end{tabular}
	\end{center}
\end{table*}


\subsection{Heat Conduction}

Asides from radiative heat transfer, some conduction is expected between the mechanically connected components of the spacecraft, thus affecting the energy radiated by it: naturally, this will tend to warm the colder elements and cool the hotter ones, until a new equilibrium situation is achieved (more rigorously, a quasi-equilibrium, since the timescale of radiation and conduction is much smaller than that of the decreasing overall power output).

Thus, one should inspect the impact of this heat transfer in the main structural elements, namely between the RTGs and the main compartment and the main compartment and the high gain antenna. Following a discussion in Ref. \cite{Francisco2011}, conduction is expected to be larger between the RTGs and the main compartment: assuming that the latter is at about $0^\circ \rm C$ (a worst case scenario, as it is warmed by the electronics to $\sim 10^\circ \rm C$ \cite{temperature1,temperature2}), while the RTGs are at $ \sim 150^\circ \rm C$, a temperature gradient of approximately $60 ~\rm K/m$ is obtained. The total cross-section of the three small diameter rods composing each truss is estimated to be of the order of $10^{-4}~\rm m^2$; these are made of aluminium, with a conductivity of approximately $240~\rm W/(m \cdot K)$. Using these figures, a total conducted power of the order of $1~\rm W$ (up to $4~\rm W$ in more conservative estimates) is achieved.

This is clearly negligible, since it is two orders of magnitude below the power of the main compartment and three orders below the RTG power. Since the temperature gradient between the main compartment to the antenna is much smaller than the one considered above, the related heat conduction is well below $\sim 1 ~\rm W$. Thus, one may safely disregard the contribution arising from conduction when computing the distribution of heat radiation.


\subsection{Thermal Force Contributions}

Using the above method, we can now compute compute the contribution of the individual components listed in Table \ref{pioneer_components}. This is achieved by integrating Eq.~(\ref{force_integration}) in three successive steps. First, the emitted radiation field given by Eq.~(\ref{lambertian}) is integrated along a closed surface, yielding the first-order effect of the emissions. Afterwards, the same radiation field is integrated along the illuminated surfaces, in order to subtract the shadow effect. Finally, the reflected radiation vector-field, given by Eqs.~(\ref{diffusive_reflection}) and (\ref{specular_reflection}), is integrated along closed surfaces, adding the contribution from reflection.

This process allows us to obtain the values for the force in terms of the emitted powers and reflection coefficients. As pointed before, the results that follow are only presented along the axis of the main antenna, since all radial components cancel-out. A positive figure indicates a sunward force.

The contribution from the front surface of the main compartment (index 4) is the easiest to compute, since there are no reflecting surfaces involved. For this reason, and the fact that this surface is perpendicular to the spacecraft's spin axis, it is effectively modeled by a single radiation source, as indicated in Table~\ref{sources}. The emitted radiation field is obtained by replacing the position and surface normal direction in Eq.~(\ref{lambertian2}). The force exerted by the radiation field on the emitting  surface is obtained by integrating Eq.~(\ref{force_integration}) along a closed surface --- in this case, chosen as a half-sphere centered at the location of the radiation source. The $z$ component of the resulting force on the emitted radiation is, as expected, given by

\begin{equation}
	F_4 = {2 \over 3} {W_{\rm front} \over c}.
\end{equation}

\begin{table*}
	\begin{center}
	\caption{Position and direction of the Lambertian source used to model each emitting surface of the Pioneer spacecraft model.}
	\vskip 0.2cm
	\begin{tabular}{c c c c}
		Emitting Surface	& Source	& Position (m)	& Surface Normal (m) \\
		\hline
		\hline
		Front wall (index 4)& 1 		& $(0,0,-0.343)$	& $(0,0,-1)$ \\
		\hline
		Lateral wall		& 1 		& $(0.572,0.2475,-0.172)$	& $(1,0,0)$ \\
		(index 1.1)			& 2 		& $(0.572,0.0825,-0.172)$	& $(1,0,0)$ \\
							& 3 		& $(0.572,-0.0825,-0.172)$	& $(1,0,0)$ \\
							& 4 		& $(0.572,-0.2475,-0.172)$	& $(1,0,0)$ \\
		\hline
		RTG					& 1 		& $(2.5,0,0)$	& $(-1,0,0)$ \\
		(index 2.1 \& 2.2)	& 2 		& $(3.1,0,0)$	& $(1,0,0)$ \\
		\hline
		Back wall			& 1 		& $(0.381,0,0)$	& $(0,0,1)$ \\
		(index 3.1)			& 2 		& $(0.191,0.33,0)$	& $(0,0,1)$ \\
							& 3 		& $(-0.193,0,33)$	& $(0,0,1)$ \\
							& 4 		& $(-0.381,0,0)$	& $(0,0,1)$ \\
							& 5 		& $(-0.191,-0.33,0)$	& $(0,0,1)$ \\
							& 6 		& $(0.191,-0.33,0)$	& $(0,0,1)$
		\label{sources}
	\end{tabular}
	\end{center}
\end{table*}

The radiation coming from the lateral walls of the main equipment compartment will illuminate the high-gain antenna (index 1.1). Due to the symmetry of the problem, and neglecting the interaction with the small far RTGs, it is only necessary to model one of the six walls. The set of Lambertian sources used for one of these walls is indicated in Table~\ref{sources}. The $z$ component of the radiation field force on the emitting surface vanishes, as the emitting surface is perpendicular to the $z$-axis.

Using Eq.~(\ref{force_integration}), but taking the integral over the illuminated portion of the antenna dish, we obtain the force exerted on the illuminated surface, which accounts for the shadow effect. This gives a $z$ component of $-0.0738(W_{\rm lat} / c)$, where $W_{\rm lat}$ is the power emitted from the lateral walls --- to be subtracted from the total force of the emitted radiation.

In what concerns the computation of diffusive reflection, Eq.~(\ref{diffusive_reflection}) allows for the computation of the reflected Poynting vector-field $\mathbf{S}_{\mathrm{rd}}(\mathbf{x},\mathbf{x}')$ due to the emitted radiation field $\mathbf{S}(\mathbf{x}')$, where $\mathbf{x}'$ is a point in the reflecting surface. The reflected radiation field is given for each point in the reflecting surface. Consequently, it must be integrated first over the reflecting surface, conveniently parameterized, giving the resulting reflected radiation field --- and then through Eq.~(\ref{force_integration}) over a closed surface, in order to compute the force resulting from the reflected radiation. The procedure for specular reflection is analogous, except that Eq.~(\ref{specular_reflection}) should be used to obtain the reflected radiation field prior to performing the integration.

Integrating the vector-field representing radiation from the lateral walls of the main compartment reflecting on the high-gain antenna, we obtain a force result of $0.0537 k_{\rm d,ant}(W_{\rm lat} / c)$ for the diffusive component and $0.0089 k_{\rm s,ant}(W_{\rm lat} / c)$ for the specular component, where $W_{\rm lat}$ is the power emitted from the referred walls and $k_{\rm d,ant}$ and $k_{\rm s,ant}$ are the diffusive and specular reflection coefficients of the main antenna, respectively. 

The result for the contribution is given by adding the emitted radiation force (zero), the shadow effect and both components of reflection, leading to

\begin{equation}
	F_{11} = {W_{\rm lat} \over c} (0.0738 + 0.0537 k_{\rm d,ant} + 0.0089 k_{\rm s,ant}).
\end{equation}

The emissions from the RTGs were modeled with two Lambertian sources, one at each base of each cylindrical shape RTG, as listed in Table~\ref{sources}. As in the case of the lateral walls, only one RTG needs to be modeled, since the effect of the radial components will be cancelled-out at each revolution of the spacecraft. It is easy to show that only the emissions from the base facing the centre of the spacecraft (source 1 of the RTG in Table~\ref{sources}) will produce a net effect on the acceleration along the $z$-axis. Emissions from the base facing outwards (source 2) are not reflected on any surface and its contribution vanishes when averaged over each revolution of the spacecraft.

Using the same procedure, the force generated by the RTG emissions is thus given in terms of the power emitted from the RTG bases facing the centre of the spacecraft $W_{\rm RTGb}$. The force resulting from reflections on the antenna (index 2.1) is given by

\begin{equation}
	F_{21} = {W_{\rm RTGb} \over c} (0.0283 + 0.0478 k_{\rm d,ant} + 0.0502 k_{\rm s,ant}),
\end{equation}

\noindent and the contribution from reflections on the lateral surfaces of the main equipment compartment is

\begin{equation}
	F_{22} = {W_{\rm RTGb} \over c} (-0.0016 + 0.0013 k_{\rm s,lat}),
\end{equation}

\noindent where $k_{\rm d,ant}$, $k_{\rm s,ant}$, $k_{\rm d,lat}$ and $k_{\rm s,lat}$ are the respective reflection coefficients.

We decided to include the computation of an additional possible contribution, which had not been previously considered in our estimates. In our previous work \cite{Bertolami2008}, it was argued that the contribution from radiation emitted from the back wall of the main compartment and reflecting in the space between this compartment and the antenna dish would be small.

In order to verify this assumption, a computation was made using the method described above. The results ultimately show that this contribution cannot be discarded after all, as it may be relevant in the final result. Considering one reflection from the antenna dish, the result in terms of the emitted power from the back wall of the main compartment $W_{\rm back}$, by

\begin{equation}
	F_{3} = {W_{\rm back} \over c} \left( -{2 \over 3} + 0.5872 + 0.5040 k_{\rm d,ant} + 0.3479 k_{\rm s,ant} \right).
\end{equation}
It should be noted that, in the preceding equation, the $-{2 \over 3}{W_{\rm back} \over c}$ is the contribution from the emitted radiation and $0.5872 {W_{\rm back} \over c}$ is the effect of the antenna's shadow. The remaining terms are the reflective contributions.

From the force computations, once the respective powers and reflection coefficients are inserted, the final result of the acceleration due to thermal dissipation mechanisms is given by

\begin{equation}
	a_{\rm th} = {F_{11} + F_{21} + F_{22} + F_{3} +F_{4} \over m_{\rm Pio}},
\end{equation}

\noindent where the mass of the spacecraft is taken at an approximate value $m_{\rm Pio} = 230~\rm{kg}$. This figure considers a total mass of $259~\rm{kg}$ at launch, including $36~\rm{kg}$ of hydrazine propellant that was partially consumed in the early stages of the mission \cite{Anderson2002}. Note that this is an approximate figure, since the actual masses for the Pioneer 10 and 11 would be slightly different due to different fuel consumptions along the mission.


\subsection{Available Power}
\label{available_power}

In this study, we chose to use the available power onboard the Pioneer spacecraft as an independent variable in the computation of the thermally induced acceleration. This choice is justified since available power is reasonably well known --- indeed, it is one of the few parameters with consistent data available throughout the operational life of the probes.

All the power on board the Pioneer probes comes from the two plutonium RTGs. It is thus easy to compute the total power available, considering the $87.74~\rm{year}$ half-life of plutonium. According to Ref. \cite{Anderson2002}, the total thermal power of the RTGs at launch was $2580~\rm{W}$. Consequently, its evolution with time will be given by

\begin{equation}
	W_{\rm tot} = 2580 \exp \left(- {t \ln 2 \over 87.72} \right)~\rm{W}
	\label{total_power}
\end{equation}

\noindent with $t$ being the time in years from launch.

The electrical power is generated by a set of thermocouples located in the RTGs. Most of this power is consumed by the various systems located in the main equipment compartment, except for a small portion used by the radio signal. A good measurement of the electrical power is available through telemetry data \cite{Toth2008}. Knowing the electrical power consumption, the remaining unused power is mostly dissipated at the RTGs themselves, via suitably designed radiating fins. Thus, we assume that the total available power is divided into two portions:

\begin{itemize}
	\item electrical power used in equipment located in the main compartment;
	\item remaining thermal power dissipated at the RTGs.
\end{itemize}

At launch, $120~\rm{W}$ of electrical power were being used in the main equipment compartment plus around $20~\rm{W}$ for the radio transmission to Earth, leaving $2420~\rm{W}$ of thermal power in the RTGs. It is also known from telemetry data that the electrical power decayed at a faster rate than thermal power, with its half-life being around $24~\rm{years}$. This would lead to an approximate time evolution of the electrical power in the equipment compartment given by

\begin{equation}
	W_{\rm equip} \approx 120 \exp \left(- {t \ln 2 \over 24} \right)~\rm{W},
	\label{elec_power}
\end{equation}

\noindent which is consistent with Fig. 11 in Ref. \cite{Toth2008}.

The baseline scenario established in this study bears the above considerations in mind and accounts for the power values extracted from the available telemetry data for the latest stages of the mission --- specifically, the reading for the twenty six years after launch (for the Pioneer 10, up to 1998). In a second stage of this study, the time evolution is taken into account, according to the reasoning developed in this section.


\section{Results and Discussion}


\subsection{Baseline Results}

In this section, a set of five scenarios are considered, while keeping the total power as $W_{\rm tot} = 2100~\rm{W}$ and the electrical power as $W_{\rm equip} = 56~\rm{W}$, leaving RTG thermal power at $W_{\rm RTG} = 2024~\rm{W}$ (assuming the power of the radio beam is still $20~\rm{W}$).

\subsubsection{Scenario 1: Lower bound, uniform temperature}

In Scenario 1, one sets a lower bound to the thermal acceleration of the Pioneer probes. The simplest possibility to consider is that each component of the spacecraft has a uniform temperature along its surfaces. In this case, the thermal power distribution will be

\begin{equation}
		W_{\rm front} = W_{\rm back} = 17.50~{\rm W}, ~~ W_{\rm lat} = 21.00~\rm{W}, ~~W_{\rm RTGb} = 143.86~\rm{W}.
\end{equation}

\noindent leading to an acceleration from thermal effects measuring $a_{\rm th} = 2.27 \times 10^{-10}~\rm{m/s^2}$.

\subsubsection{Scenario 2: Higher emissions from louvers}
Scenario 2 assumes that the front wall, including the louvers, is responsible for $W_{\rm front} = 40~\rm{W}$ (that is, $70\%$ of $56~\rm{W}$) of emission. This leaves the lateral walls with  $W_{\rm lat} = 8.73~\rm{W}$ and the back wall with $W_{\rm back} = 7.27~\rm{W}$. The thermally induced acceleration in these conditions is $a_{\rm th} = 4.43 \times 10^{-10}~\rm{m/s^2}$. This scenario is motivated by one essential feature: the louvers located in the front wall of the main equipment compartment were designed to act as a temperature controlling element, closing or opening through the action of a bi-metallic spring. Still, even when closed, the louvers are not covered by the Multi-Layer Insulation (MLI) which shields the equipment compartment. It is then reasonable to assume that, regardless of their position, the louvers radiate a large share of the equipment power. A similar argument is presented in Ref. \cite{Scheffer2003}.

\subsubsection{Scenario 3: Diffusive reflection in antenna}
In Scenario 3, one includes the contribution from reflections. The simplest way to achieve this is to include only the diffusive component. We will consider a diffusive reflection coefficient of $k_{\rm d,ant}=0.8$, which would be a typical value for aluminum, used in the antenna dish. This yields a result for the acceleration of $a_{\rm th} = 5.71 \times 10^{-10}~\rm{m/s^2}$.

\subsubsection{Scenario 4: Diffusive and specular reflection}
Scenario 4 is a variation that considers the fact that illuminated surface of the high-gain antenna is made of bare aluminum \cite{Turyshev2008}. This will make reflection from it mainly diffusive, but with a small specular highlight, as is typical of any unpolished flat surface. We thus consider a reflection from the antenna dish that maintains a total coefficient of $80\%$, but divided in diffusive and specular components --- respectively, $k_{\rm d,ant}=0.6$ and $k_{\rm s,ant}=0.2$. Furthermore, we assume a specular reflection from the MLI covering the main equipment compartment of $k_{\rm s,lat}=0.4$. The result from this scenario is not significantly different from Scenario 3, yielding $a_{\rm th} = 5.69 \times 10^{-10}~\rm{m/s^2}$.

\subsubsection{Scenario 5: Upper bound	}
In Scenario 5, in order to obtain an upper bound for the static baseline, one assumes that all the emissions of the main equipment compartment come from the louvers and a $10\%$ higher power from the RTG base, that is,

\begin{equation}
		W_{\rm front} = 56~{\rm W}, ~~ W_{\rm back} = W_{\rm lat} = 0~\rm{W}, W_{\rm RTGb} = 158.24~\rm{W}.
\end{equation}

\noindent Maintaining $k_{\rm d,ant}=0.8$ as in Scenario 3, the upper bound for the thermal acceleration in the late stage of the mission is bound to be $a_{\rm th} = 6.71 \times 10^{-10}~\rm{m/s^2}$.

The results of all the considered scenarios are summarized in Table \ref{baseline_results}.

\begin{table*}[ht]
	\begin{center}
	\caption{Pioneer thermal acceleration results for baseline scenarios.}
	\vskip 0.2cm
	\begin{tabular}{c | c c c c | c c c | c}
	
		 Scenario	& $W_{\rm RTGb}$	& $W_{\rm front}$	& $W_{\rm lat}$	& $W_{\rm back}$	& $k_{\rm d,ant}$	& $k_{\rm s,ant}$	& $k_{\rm s,lat}$	& $a_{\rm th}$ \\
		 				& $(\rm{W})$		& $(\rm{W})$		& $(\rm{W})$	& $(\rm{W})$		&	&	&	& $(10^{-10}~\rm{m/s^2})$ \\
		\hline
		\hline
		1	&$143.86$	&$17.5$	&$21$	&$17.5$	&$0$ 	&$0$ 	&$0$ & $2.27$ \\
		2	&$143.86$	&$40$	&$8.73$	&$7.27$	&$0$ 	&$0$ 	&$0$ & $4.43$ \\
		3	&$143.86$	&$40$	&$8.73$	&$7.27$	&$0.8$ 	&$0$ 	&$0$ & $5.71$ \\
		4	&$143.86$	&$40$	&$8.73$	&$7.27$	&$0.6$ 	&$0.2$ 	&$0.4$ & $5.69$ \\
		5	&$158.24$	&$56$	&$0$	&$0$	&$0.8$ 	&$0$ 	&$0.4$ & $6.71$ 
	
		\label{baseline_results}
	\end{tabular}
	\end{center}
\end{table*}

With these baseline scenarios, we shall proceed with a static parametric analysis of the involved parameters in order to obtain a result and an error bar for the 1998 static figures.


\subsection{Parametric Analysis}

As outlined above, the first step is to perform a static parametric analysis, in an attempt to establish an estimate for the thermal acceleration at an instant 26 years after launch. The analysis is performed using a classic Monte-Carlo method, where a probability distribution is assigned to each variable and random values are then generated. A distribution of the final result ({\it i.e.} the acceleration) is then obtained.

The parameters that come into play in this setup are the power emitted from each surface, $W_{\rm RTGb}$, $W_{\rm front}$, $W_{\rm lat}$, $W_{\rm back}$, and the reflection coefficients $k_{\rm d,ant}$, $k_{\rm s,ant}$ and $k_{\rm s,lat}$.

A quick analysis of Table \ref{baseline_results} allows us to draw some qualitative conclusions: for example, the amount of power emitted from the front wall $W_{\rm front}$ has a decisive influence in the final result. In contrast, the relevance of the specular reflection coefficient of the lateral wall $k_{\rm s,lat}$ is almost negligible.

For the static analysis at $t=26~\mathrm{years}$, Scenario 4 is taken as a reference, since it is the one more solidly based on physical arguments.

The power emitted by the RTG bases facing the main compartment $W_{\rm RTGb}$ is generated from a normal distribution with the mean value of $143.86~\rm{W}$ and a standard deviation of $25\%$ of this value. This allows for a significantly larger deviation than that considered in the top-bound scenario (Scenario 5), which had only a $10\%$ increase in the power of this surface. The purpose is to account for unanticipated anisotropies in the temperature distribution of the RTGs.

In the case of the main equipment compartment, the focus is on the power emitted by the louvers located in the front wall. The selected distribution for the parameter $W_{\rm front}$ is also normal, with the mean value at $40~\rm{W}$ (also corresponding to Scenario 4). We set the standard deviation at $7.5~\rm{W}$, so that the 95\% probability interval ($2 \sigma$) for the value of $W_{\rm front}$ is below the top figure of $56~\rm{W}$, which corresponds to the totality of the equipment power being dissipated in the front wall. For the remaining surfaces of the equipment compartment, the power is computed at each instance so that the total power of the equipment is conserved at $56~\rm{W}$.

Concerning the reflection coefficients for the antenna, we use uniform distributions in the intervals $[0.6,0.8]$ for $k_{\rm d,ant}$ and $[0,0.2]$ for $k_{\rm s,ant}$, while imposing the condition $k_{\rm d,ant}+k_{\rm s,ant}=0.8$, since this is a typical value for aluminum in infrared wavelengths. We also expect the specular component to be small, since the surface is not polished. Furthermore, if we allow for the possibility of surface degradation with time during the mission, the specular component would suffer a progressive reduction in favor of the diffusive component, a possibility that this analysis takes into account.

We performed $10^4$ Monte Carlo iterations, which easily ensures the convergence of the result. The thermal acceleration estimate yielded by the simulation for an instant $26~\mathrm{years}$ after launch, with a $95\%$ probability, is

\begin{equation}
	a_{\rm{th}}(t=26)=(5.8 \pm 1.3) \times 10^{-10} ~ \rm{m/s^2}.
\end{equation}

\begin{figure}
	\begin{center}
	\epsfxsize=0.9\columnwidth
	\epsffile{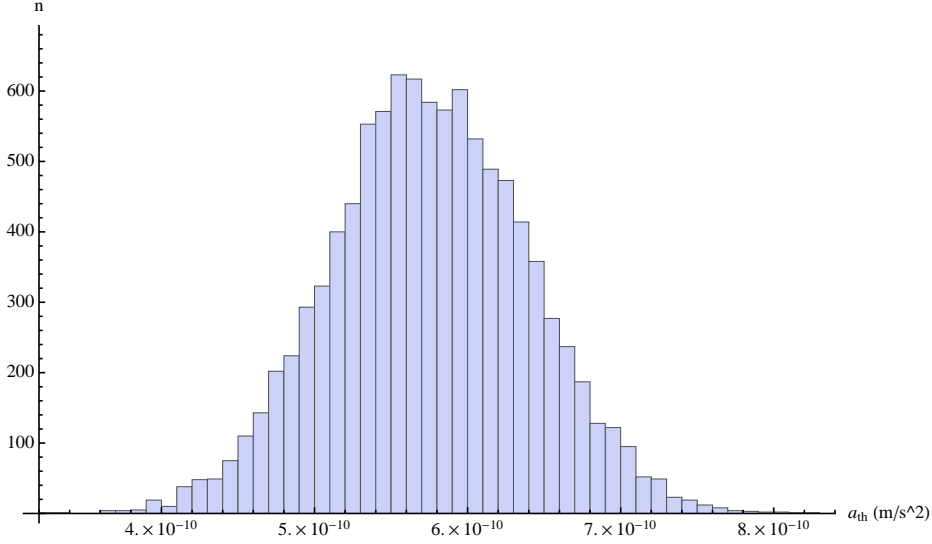}
	\caption{Histogram for the distribution resulting from the Monte Carlo simulation with 10000 iterations for the thermal acceleration of the spacecraft at $t=26~\mathrm{years}$ after launch.}
	\label{static_hist}
	\end{center}
\end{figure}

\noindent This result is extracted from the approximately normal distribution shown in Fig.~\ref{static_hist}; the conformity of the results to a normal distribution was confirmed by a Shapiro-Wilk normalcy test with a $p$-value $\sim 0$.

These results account for between $44\%$ and $96\%$ of the reported value $a_{\rm{Pio}}=(8.74 \pm 1.33) \times 10^{-10}~\rm{m/s^2}$ (which, we recall, was obtained under the hypothesis of a constant acceleration) --- thus giving a strong indication of the preponderant contribution of thermal effects to the Pioneer anomaly.


\subsection{Time Evolution}

The final step of this study is to perform an analysis of the expected time evolution of the thermal acceleration affecting the Pioneer 10 and 11 spacecraft.

An immediate estimate can be obtained by extrapolating the static results with the available time evolution of electric power, using Eqs. (\ref{total_power}) and (\ref{elec_power}). Results are shown as the dotted line in Fig.~\ref{time_evolution}, with the approximate exponential decay of the available power translated into a similar trend in the evolution of the thermally induced acceleration.

This extrapolation, however, does not account for the possibility that some parameters may change with time --- namely, the power distribution throughout the different surfaces or their reflection coefficients. This could be accounted by a simulation of the full span of the missions ({\it i.e.} a large number of consecutive simulations), with a specific prescription for the variability of these parameters.

Such task would prove too lengthy, and no significant physical insight would be gained. Hence, we have preferred a somewhat simpler approach: for a better grasp of the possibility discussed above, we apply the Monte-Carlo static analysis to only two earlier moments of the mission. Each simulation produces a central value, with top and lower bounds; these are then fitted to an exponential trend, thus obtaining an estimate of the time evolution of the thermally induced acceleration.

The selected instant for the earliest static analysis was at $t=8 ~\mathrm{years}$ after launch, corresponding to the 1980 values for the Pioneer 10. This corresponds to the time at which the effect of the solar radiation pressure dropped below $5 \times 10^{-10}~\rm{m/s^2}$ \cite{Anderson2002}.

\begin{figure}
	\begin{center}
	\epsfxsize=0.9\columnwidth
	\epsffile{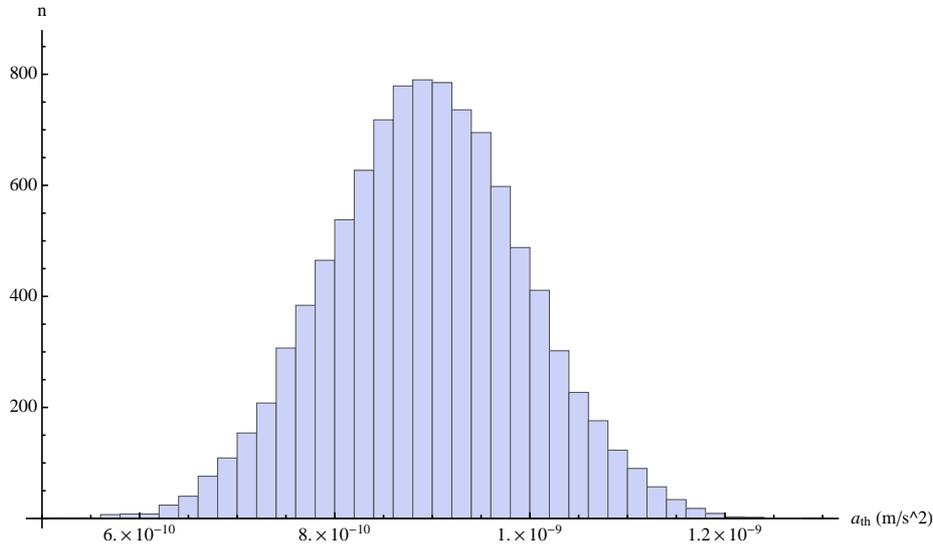}
	\caption{Histogram for the distribution resulting from the Monte Carlo simulation with 10000 iterations for the thermal acceleration of the spacecraft at $t=8~\mathrm{years}$ after launch.}
	\label{static_hist_early}
	\end{center}
\end{figure}

\begin{figure}
	\begin{center}
	\epsfxsize=0.8\columnwidth
	\epsffile{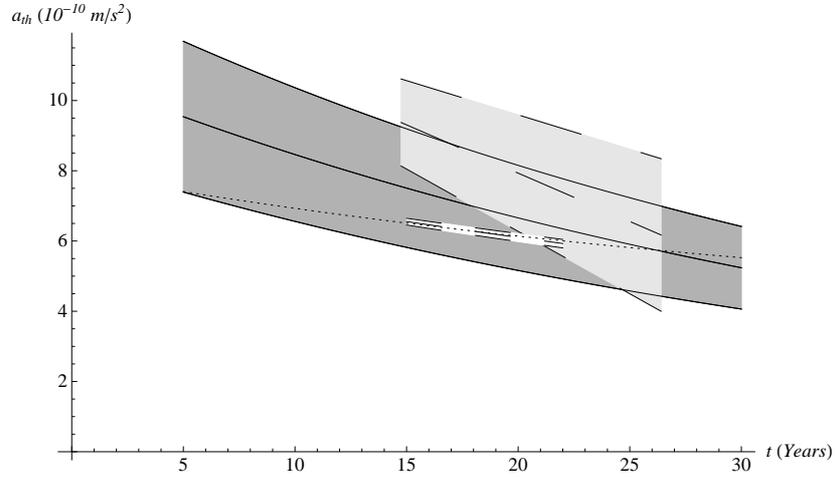}
	\caption{Results for the time evolution of the thermal acceleration on the Pioneer spacecraft compared with results based on two data analyses with non-constant solutions for the anomalous acceleration. The dotted line is the time extrapolation of the static analysis of the thermal acceleration and the dark grey area correspond to a $95\%$ probability for the thermal acceleration in the time evolution analysis. For comparison, the light grey area is based on results from the data analysis in Refs.~\cite{Toth2009} and \cite{Markwardt2002}, respectively.}
	\label{time_evolution}
	\end{center}
\end{figure}

This analysis is made in a similar fashion as the one presented in the previous subsection, but using the 1980 available power values as a base for the choice of the distributions. The thermal acceleration is, in this case,

\begin{equation}
	a_{\rm{th}}(t=8)=(8.9 \pm 2) \times 10^{-10} ~ \rm{m/s^2},
\end{equation}

\noindent corresponding to the same $95\%$ probability in the approximately normal distribution in Fig.~\ref{static_hist_early}.

The values obtained here for this earlier stage of the mission bear a close match to those of the assumed constant anomalous acceleration.

The third static analysis was performed at a time $t=17~\mathrm{years}$, halfway between the other two. The estimate in this case is, for a $95\%$ probability,

\begin{equation}
	a_{\rm{th}}(t=17)=(7.1 \pm 1.6) \times 10^{-10} ~ \rm{m/s^2}.
\end{equation}

Using the three static estimates presented above, it is now possible to produce a time evolution based on a fit to an exponential decay. This is performed for the mean value, top-bound and lower-bound of the acceleration, always based on a $95\%$ probability degree.

The curve fit for the mean, upper and lower values of the thermal acceleration reads

\begin{equation}
	a_{\rm th} =  [(1.07 \pm 0.24) \times 10^{-9}] \exp (-0.0240t)~\rm{m/s^2},
\end{equation}

\noindent with $t$ giving the time after launch in $\mathrm{years}$.

The time evolution resulting from any of these scenarios corresponds to a decay with a half-life of approximately $60~\mathrm{years}$, related to the nuclear decay of the plutonium in the RTGs and the faster decay rate of the electrical power, already discussed in Section~\ref{available_power}. The graphic representation of the band of values predicted by our model is shown in Fig~\ref{time_evolution} (dark grey region) and compared with the values indicated by non-constant results for the anomalous acceleration in Refs.~\cite{Toth2009,Markwardt2002} (light grey region).

\section{Conclusion and Outlook}

In this study we have established a new method to model reflections of the Pioneer spacecraft thermal radiation with an increased accuracy, while maintaining the desired simplicity and computational speed of the approach previously proposed \cite{Bertolami2008}. This new tool allows for a successful modeling of the most important features of the Pioneer spacecraft concerning thermal effects and its impact on the resulting acceleration.

The developed method, based on Phong shading, provides results that generally confirm those previously obtained in Refs.~\cite{Bertolami2008,Bertolami2010}: the acceleration arising from thermal radiation effects has a similar order of magnitude to the constant anomalous acceleration reported in Ref. \cite{Anderson2002}. We believe that the chosen approach is most adequate for the study of this particular problem, taking into account all its specific characteristics. Moreover, this Phong shading method is well suited for future studies of radiation momentum transfer in other spacecraft.

The main difficulty in dealing with this problem has always been the lack of sufficient and reliable information for a detailed engineering modeling of the spacecraft, which justified a large number of reasonable hypotheses. We have achieved to overcome this caveat through a parametric analysis that takes into account a wide range of different scenarios. This strategy allows us to present a range of probable values for the thermal effects, which appears to be compatible with the signature of the Pioneer anomalous acceleration.

With the results presented here it becomes increasingly apparent that, unless new data arises, the puzzle of the anomalous acceleration of the Pioneer probes can finally be put to rest.


\section*{Acknowledgments}

\samepage
\noindent

The work of FF is sponsored by the FCT -- Funda\c{c}\~{a}o para a Ci\^{e}ncia e Tecnologia (Portuguese Agency), under the grant BD 66189/2009.


\bibliographystyle{model1-num-names}
\bibliography{phong_shading}

\end{document}